\documentclass[12pt]{iopart}

\usepackage[english]{babel}
\usepackage{iopams}
\usepackage{revsymb}
\usepackage{amsfonts}
\usepackage{amssymb}
\usepackage{graphicx}
\expandafter\let\csname equation*\endcsname\relax
\expandafter\let\csname endequation*\endcsname\relax
\usepackage{amsmath}
\usepackage{color}

\newcommand{\se}{\textsc{S.}~}
\newcommand{\sse}{\textsc{Subs.}~}
\newcommand{\eq}{\textsc{Eq.}~}
\newcommand{\eqs}{\textsc{Eqs.}~}

\newcommand{\fig}{\textsc{Fig.}~}
\newcommand{\figs}{\textsc{Figs.}~}
\newcommand{\tper}{t_{\rm p}}

\begin{document}
\title[]{Exactly solvable model of stochastic heat engine: Optimization of power, its fluctuations and efficiency}

\author{Viktor Holubec}
\address{
Charles University in Prague, Faculty of Mathematics and Physics, Department of Macromolecular Physics, V Hole{\v s}ovi{\v c}k{\'a}ch 2 ,180~00~Praha, Czech Republic}
\eads{\mailto{viktor.holubec@gmail.com}}
\date{\today}
\begin{abstract}
We investigate a stochastic heat engine based on an over-damped particle diffusing on the positive real axis in an externally driven time-periodic log-harmonic potential. The periodic driving is composed of two isothermal and two adiabatic branches. Within our specific setting we verify the recent universal results regarding efficiency at maximum power and discuss  properties of the optimal protocol. Namely, we show that for certain fixed parameters the optimal protocol maximizes not only the output power but also the efficiency. Moreover, we calculate the variance of the output work and discuss the possibility to minimize fluctuations of the output power. 
\end{abstract}
\maketitle
\section{Introduction}

Stochastic thermodynamics is an advancing field with many applications to small systems 
of current interest \cite{Jarzynski1997, Carberry2004, Seifert2008, Hanggi2009, Speck2011, Bressloff2013}. 
One of the hot topics are stochastic heat engines \cite{Seifert2012}. The engines so small that their dynamics is significantly influenced by thermal fluctuations of the surrounding environment and thus both their state and its functionals such as work and heat become stochastic. Common thermodynamic quantities are then obtained as averages of their fluctuating counterparts. Models studied in literature can be roughly classified according to the dynamical laws involved. In the case of the classical stochastic heat engines, the state space can either be discrete or continuous (c.f., for example, \cite{Sekimoto2000, Schmiedl2008, Chvosta2010, VandenBroeck2004} and the references in the thorough review \cite{Seifert2012}). Examples of the quantum heat engines are studied, e.g., in \cite{Allahverdyan2008, Henrich2007, Abah2012, Rossnagel2014}. 
One of the most interesting theoretical results obtained in the field are recent elegant formulas for efficiency of heat engines working at maximum power (EMP)  \cite{Schmiedl2008, Esposito2009, Esposito2010b, Zhan-Chun2012}.

In the present paper, we focus on a classical stochastic heat engine based on non-interacting Brownian particles diffusing in a periodically driven potential. Such heat engines usually operate on time-scales smaller than intrinsic relaxation times of their working medium and hence far from quasi-static conditions, when the work fluctuations are even more pronounced (one example is \fig2 in the amazing experimental study \cite{Blickle2011}, where the measured work fluctuations are depicted). Therefore the common analysis of the stochastic heat engines, which does not consider \emph{fluctuations} (one exception is the study \cite{Chvosta2010}), may not be always sufficient. Strong general results concerning EMP for a wide class of such engines were obtained by Schmiedl and Seifert \cite{Schmiedl2008}. These authors also demonstrated their findings on a specific exactly solvable model describing an engine driven by a \emph{harmonic potential}. To the best of our knowledge, no more exactly solvable models which would demonstrate the Schmiedl's and Seifert's results were published. 

We give an exactly solvable example of a stochastic heat engine based on a particle diffusing in the \emph{log-harmonic potential} \cite{Giampaoli1999,Strier2000,Ryabov2013}. The periodic driving is composed of two isothermal and two adiabatic branches. Within our specific setting we verify the universal results regarding EMP obtained in \cite{Schmiedl2008, Esposito2009, Esposito2010b} and discuss  properties of the optimal protocol. Namely, we show that for certain fixed parameters the optimal protocol maximizes also the \emph{efficiency}. Moreover, we calculate the \emph{variance of the output work} and discuss the possibility to \emph{minimize fluctuations} of the output power. 

The paper is organized as follows: in \se\ref{sec:model} we 
introduce the working medium of the engine and present formulas describing its dynamics. Further, in \se\ref{sec:thermodynamics}, we specify the thermodynamic quantities used in our analysis. 
\se\ref{sec:diagrams} contains the discussion concerning 
the possibility to depict the engine working cycle
in a stochastic thermodynamics analogue of the standard $p$-$V$ diagrams. 
In \se\ref{sec:driving} we introduce three specific examples 
of the driving. First, we derive the protocol which maximizes the
output power. Second, we define the driving which yields the smallest 
power fluctuations from the three protocols. These two protocols 
inevitably incorporate two isothermal and two adiabatic branches. 
The third protocol is different, it can be composed of two isotherms
only. The discussion and illustrations of the obtained results are postponed to \se\ref{sec:discussion}, where we compare performance of these three protocols. 
\section{Engine and its working cycle}
\label{sec:model}

Consider an over-damped Brownian particle diffusing on the interval $[0,\infty]$ with the reflecting boundary at $x=0$. Due to the thermal fluctuations the position of the particle forms a stochastic process $x(t)$. It is, therefore, necessary to describe the state of the system using the probability density $\rho(x, t)$ to find the particle in a position $x$ at a time $t$. Assume that the diffusion occurs in aqueous environment with a time-periodic temperature $T(t)$ and that 
the Brownian particle is subjected to a time-periodic log-harmonic external potential
\cite{Giampaoli1999,Strier2000,Ryabov2013}
\begin{equation}
{\cal V}(x,t) = - g(t) \log x + \frac{k(t)}{2}x^2\,\,,\quad- g(t) < k_{\rm B} T(t)\,\,,\quad k(t) > 0\,\,,
\label{eq:potential}
\end{equation}
where $k_{\rm B}$ stands for the Boltzmann constant. Then the transition probability density for the position of the particle, $R(x,t\,|\,x',t')$, can be described by the Fokker-Planck equation \cite{Risken1985}
\begin{equation}
\frac{\partial}{\partial t} R(x,t\,|\,x',t')
= \left\{ 
D(t) \frac{\partial^2}{\partial x^2} - \frac{\partial}{\partial x} \left[ \frac{g(t)}{x} - k(t)\,x \right] \right\} R(x,t\,|\,x',t')
\label{eq:fokker_planck}
\end{equation}
with the initial condition $R(x,t'\,|\,x',t') = \delta(x - x')$. Note that due to the time-dependence of the temperature the diffusion coefficient $D(t) = k_{\rm B} T(t)/\gamma$ also depends on time ($\gamma$ denotes the friction coefficient). Starting from an arbitrary probability density $\rho(x', t')$ at a time $t'$ the Green's function $R(x,t\,|\,x',t')$ yields the state of the system in any subsequent time via single integration
\begin{equation}
\rho(x,t) = \int_{0}^{\infty}{\rm d}x'\, \rho(x',t')R(x,t\,|\,x',t')\,\,.
\label{eq:PDF_general}
\end{equation}

Under the periodic driving, ${\cal V}(x,t) = {\cal V}(x,t + t_{\rm p})$, $T(t) = T(t + t_{\rm p})$, the state of the system eventually, after a transient period, becomes also periodic. The system attains a limit cycle -- the engine working cycle. The conditions on the functions $g(t)$ and $k(t)$ in \eq(\ref{eq:potential})
secure that a non-trivial time-asymptotic solution of \eq(\ref{eq:fokker_planck}) exists. 
For example, assume that both the strength of the logarithmic part of the potential $g$ and the temperature $T$ are time-independent. Then, in the case of $- g \geq k_{\rm B} T$ and $k(t) > 0$, the particle eventually attains the coordinate $x = 0$ with unit probability, i.e. $\lim_{t \rightarrow \infty} \rho(x,t) = \delta(x)$ for an arbitrary $\rho(x', t')$.

Let us denote the state of the engine during the limit cycle as $p(x,t)$. If the system starts from the state $p(x,0)$ at the time $t=0$, this state is revisited after each period of the driving, $t_{\rm p}$. From \eq(\ref{eq:PDF_general}) then follows:
\begin{equation}
\label{eq:EigenvalueProblem}
p(x,0)=\int_{0}^{\infty}{\rm d}x'\,
R(x,t_{\rm p}\,|\,x',0')\,
p(x',0)\,\,.
\end{equation}

In order to solve this integral equation for $p(x,0)$, we need to find the specific form of the Green's function $R(x,t\,|\,x',t')$ first. Known solutions of \eq(\ref{eq:fokker_planck}) are available only for the case of time-independent diffusion constant and strength of the logarithmic part of the potential \cite{Ryabov2013}, i.e. for $T(t) = T$, $g(t) = g$. Therefore, in solving \eq(\ref{eq:fokker_planck}) we  proceed in two steps:
1) we find the Green's function for a generic isothermal driving with constant $g$, 2) we use the Markov property of the underlying stochastic process and compose the solution of \eq(\ref{eq:fokker_planck}) for a periodic driving from the generic solutions. To this end we use the Chapman-Kolmogorov condition
\begin{equation}
\label{eq:RchapmankolmogorovContinuous}
R(x,t\,|\,x',t')=\int_{0}^{\infty}{\rm d}x''\,
R(x,t\,|\,x'',t'')
R(x'',t''\,|\,x',t')\,\,,
\end{equation}
which is fulfilled by the Green's function for any intermediate time $t''$. Using the described procedure one can find solution of \eq(\ref{eq:fokker_planck}) for any piecewise constant functions $D(t)$ and $g(t)$, although in the rest of the paper we consider only two isothermal branches. In the following we set $\gamma = k_{\rm B} = 1$. 

\subsection{Generic case -- isothermal process}

Consider an isothermal process with the driving
\begin{equation}
{\cal V}(x,t) = - g \log x + \frac{k(t)}{2}x^2\,\,.
\label{eq:energy_continuousmotor_generic}
\end{equation}
In this generic case [$T(t) = T$, $g(t) = g$], the solution of \eq({\ref{eq:fokker_planck}}) reads \cite{Ryabov2013}
\begin{equation}
R_{\rm g}(x,t\,|\,x',t') = \frac{x' {\rm e}^{\frac{\nu+2}{2}a(t,t')}}{2\, b(T;t,t')}\left(\frac{x}{x'}\right)^{\nu+1}
\exp\left[-\frac{x^2{\rm e}^{a(t,t')}+(x')^2}{4\,b(T;t,t')} \right] \,\,{\rm I}_{\nu}~\!\!\left[\frac{x x'{\rm e}^{\frac{1}{2}a(t,t')}}{2\,b(T;t,t')} \right]\,\,,
\label{eq:green_function}
\end{equation}
where ${\rm I}_{\nu}~\!\!(x)$ stands for the modified Bessel function of the first kind \cite{Abramowitz1964},
\begin{equation}
\nu = \frac{1}{2} \left(\frac{g}{T} - 1\right)\,\,,\qquad \nu > - 1 
\label{eq:nu}
\end{equation}
and
\begin{equation}
\label{eq:auxiliary_bs}
\begin{split}
a(t,t') &= 2\int_{t'}^{t}{\rm d}t''\,k(t'')\,\,,
\\
b(T;t,t') &= T\int_{t'}^{t}{\rm d}t''\,\exp\left[ 2\int_{t'}^{t''}{\rm d}t'''\,k(t''') \right]\,\,.
\end{split}
\end{equation}
Note that the parameter $g$ enters the generic solution 
(\ref{eq:green_function}) only through the dimensionless
combination $\nu$.


\subsection{Engine working cycle}
\label{sec:operational_cycle}

\begin{figure}
	\centering
		\includegraphics[width=1.0\linewidth]{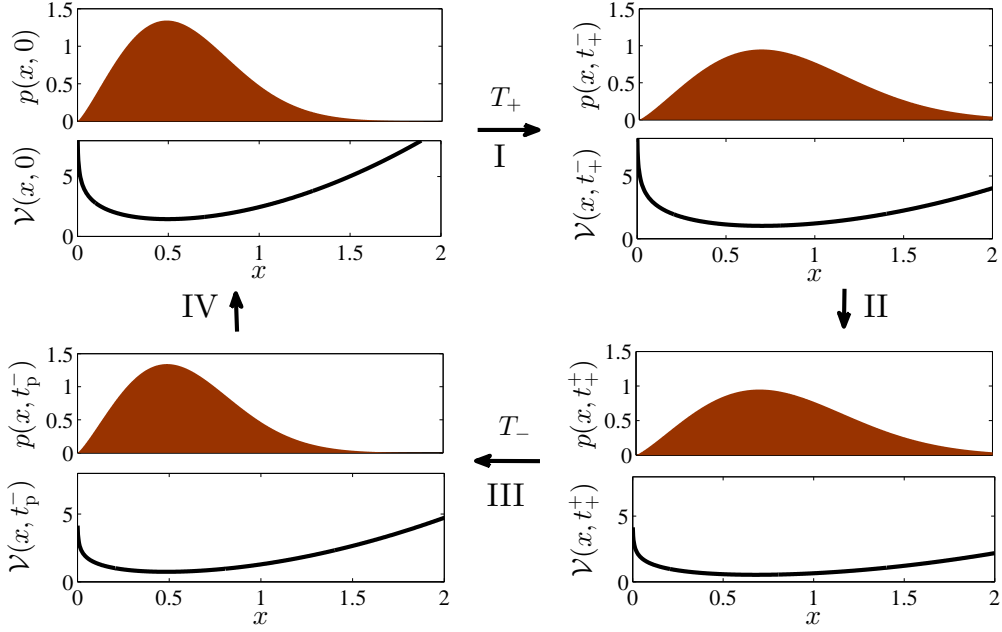}		
		\caption{Scheme of the working cycle of the engine. The particle distribution (\ref{eq:cycle_dist}) is depicted by the filled curve. The solid line represents the potential energy (\ref{eq:potential}). The logarithmic part of the potential energy is in this case repulsive.}
		\label{fig:cycle_scheme}
\end{figure}
Let the periodic driving (\ref{eq:potential})
consists of two isothermal and two adiabatic branches as it is depicted in \fig\ref{fig:cycle_scheme}. 
To completely specify the
protocol it is enough to describe the concrete form of the driving during
the isotherms. To this end, we assume that the potential energy ${\cal V}(x,t)$ has the generic form 
(\ref{eq:energy_continuousmotor_generic}), i.e. we have
\begin{equation}
{\cal V}(x,t) = \left\{ 
  \begin{array}{l l}
    \,- g_+ \log x +\displaystyle \frac{k(t)}{2}x^2\,\,, & \qquad t \in [0,t_+^-]\\
    \\
    \,- g_- \log x +\displaystyle \frac{k(t)}{2}x^2\,\,, & \qquad t \in [t_+^+,t_{\rm p}^-]
  \end{array} \right.
  \,\,.
\label{eq:cycle_energies}
\end{equation}
During the first adiabatic branch the potential changes from ${\cal V}(x,t_+^-)$
to ${\cal V}(x,t_+^+)$. During the second adiabatic branch the potential returns to
its initial value, i.e. it changes from  ${\cal V}(x,t_{\rm p}^-)$ to 
${\cal V}(x,t_{\rm p}) = {\cal V}(x,0)$. This pattern is periodically repeated, 
the period being $t_{\rm p}$.  The durations of the adiabatic branches $(t_+^+ - t_+^-)$
and $(t_{\rm p} - t_{\rm p}^-)$ are considered as infinitesimally short as compared
to the durations of the isotherms $t_+ = t_+^-$ and $t_- = (t_{\rm p}^- - t_+^+)$.
During the first (the second) isotherm the temperature $T$ assumes the value $T_+$
($T_-$). For the sake of mathematical simplicity we choose the parameters 
$g_{\pm}$ proportional to the corresponding reservoir temperatures. Specifically we take
$g_{\pm} = (2\nu+1)T_{\pm}$~\footnote{
It turns out that for general parameters $g_{\pm}$
the Green's function (\ref{eq:green_function_cycle}) is
given by a sum of Gauss hypergeometric functions \cite{Prudnikov1986} and the integral 
\eq(\ref{eq:EigenvalueProblem}) becomes quite complicated.
}. This
setting allows us to investigate the cases when the logarithmic part of the potential 
${\cal V}(x,t)$ is during the whole cycle either repulsive or attractive.
The model with general $g_{\pm}$ would describe also the situation when the two constants have different
signs and thus the logarithmic part of the potential (\ref{eq:cycle_energies}) would be during the first isotherm repulsive (attractive) and vice versa during the second one. 

The resulting driving for $\nu=-1/2$ (logarithmic part of the potential vanishes) coincides with that discussed in 
\cite{Schmiedl2008}.
The fact that the protocol contains adiabatic branches II and IV (see \fig\ref{fig:cycle_scheme}), where the potential 
and the temperature change infinitely fast while the particle distribution remains 
unchanged, may look artificial. Nevertheless, such protocol can be realized with sufficient accuracy in experiments  
\cite{Blickle2011}. 
Further, we focus on the characterization of the limit cycle, which the
engine approaches at long times after a transient period.

We start from the generic solution (\ref{eq:green_function}) of 
the Fokker-Planck equation (\ref{eq:fokker_planck}). Owing to the 
Chapman-Kolmogorov condition (\ref{eq:RchapmankolmogorovContinuous}), the Green's function 
within the cycle still assumes the form (\ref{eq:green_function}). 
Specifically it reads
\begin{equation}
R(x,t\,|\,x',t') = \frac{x' {\rm e}^{\frac{\nu+2}{2}a(t,t')}}{2 b(t,t')}
\left(\frac{x}{x'}\right)^{\nu+1}
\exp\left[-\frac{x^2{\rm e}^{a(t,t')}+(x')^2}{4b(t,t')} \right] 
\,\,{\rm I}_{\nu}~\!\!\left[\frac{x x'{\rm e}^{\frac{1}{2}a(t,t')}}{2 b(t,t')} \right]\,\,,
\label{eq:green_function_cycle}
\end{equation}
where the function $b(t,t')$ is given by
\begin{equation}
 \label{eq:auxiliary_b_cycle}
\rule[-1ex]{-1.5em}{6ex}
b(t,t') =
\left\{ \begin{array}{lll}
b(T_+;t,t')\,\,, & t',t\in[0,t_+^{+}]\,\,,\\[1ex]
\displaystyle
b(T_+;t_+,t') + {\rm e}^{a(t_+,t')}\,b(T_-;t,t_+)\,\,,
& t'\in[0,t_+^{+}] \,\, \wedge \,\, t\in[t_+^{+},\tper]\,\,,\\[1ex]
b(T_-;t,t')\,\,, & t',t\in[t_+^{+},t_{\rm p}]\,\,,
\end{array} \right.
\end{equation}
and the functions $a(t,t')$ and $b(T;t,t')$ are defined in
\eq(\ref{eq:auxiliary_bs}).

The periodic state of the system during the limit cycle 
is determined by the solution of the integral 
equation (\ref{eq:EigenvalueProblem}). 
For our specific setting it can be solved 
analytically. The result,
\begin{equation}
p(x,t) =  \frac{1}{\Gamma~\!\!(\nu+1)}\left[\frac{1}{f(t)}\right]^{\nu+1}
\left(\frac{x}{2}\right)^{2\nu+1}\exp\left[-\frac{x^2}{4f(t)}\right]\,\,,
\label{eq:cycle_dist}
\end{equation}
was already mentioned in \cite{Ryabov2013} as a time-asymptotic distribution 
for a periodically driven system in contact with a single heat bath at a constant temperature. Here, $\Gamma~\!\!(x)$ denotes the Gamma function and the function 
$f(t) = \left< [x(t)]^2 \right>/(4\nu + 4)$ determines the width 
of the distribution (\ref{eq:cycle_dist}). It reads
\begin{equation}
f(t) = \left\{ 
  \begin{array}{l l}
    \,[f_0 + b(t,0)]\exp[-a(t,0)]\,\,, & \qquad t \in [0,t_+]\\
    \,[f_1 + b(t,t_+)]\exp[-a(t,t_+)]\,\,, & \qquad t \in [t_+,t_{\rm p}]
  \end{array} \right.
  \,\,,
\label{eq:function_f}
\end{equation}
where $f_0 = f(0) = f(t_{\rm p}) =
 \left\{b(t_+,0)+b(t_{\rm p},t_+)\exp[a(t_+,0)]\right\}/h$,
$f_1 = f(t_+^-) = f(t_+^+) = \left\{b(t_{\rm p},t_+)+b(t_+,0)\exp[a(t_{\rm p},t_+)]\right\}/h$, 
and $h=\exp[a(t_{\rm p},0)]-1$.

\eqs(\ref{eq:green_function_cycle})-(\ref{eq:function_f}) represent the main result of the present section. Note that both the Green's function $R(x,t\,|\,x',t')$ and the probability density $p(x,t)$ are continuous regardless the discontinuities in the driving. In the next section we show how these functions render the whole thermodynamic description of the engine. The functions  (\ref{eq:green_function_cycle})-(\ref{eq:function_f}) and thus also the specific form of the limit cycle and the thermodynamic variables are determined solely by the externally controlled parameters $\nu$, $T_{\pm}$, $t_{\pm}$, and by the function $k(t)$.



\section{Thermodynamic quantities}
\label{sec:thermodynamics}

\subsection{Mean values}
\label{sbs:mean_values}
The probability distribution $p(x,t)$ renders the energetics of the engine 
in terms of mean values. We use the standard definitions of thermodynamic quantities as introduced for example in \cite{Seifert2008}.
The thermodynamic work done by 
the engine during the time interval $[t,t']$, $W_{\rm out}(t,t') = -\int_{t'}^{t}{\rm d}t''\,\int_{0}^{\infty}{\rm d}x\, p(x,t'')\, {\rm d}{\cal{V}}(x,t'')/{\rm d}t''$, reads 
\begin{equation}
W_{\rm out}(t,t') = \int_{t'}^{t} {\rm d}t''\, \left<\log{x(t'')}\right>\,\frac{{\rm d}g(t'')}{{\rm d}t''} 
- \int_{t'}^{t} {\rm d}t''\, \frac{1}{2} \left<[x(t'')]^2\right> \frac{{\rm d}k(t'')}{{\rm d}t''}\,\,,
\label{eq:mean_work}
\end{equation}
where the two averages are given by
\begin{align}
\left< \log x(t)\right> &= \int_0^{\infty}{\rm d}x\,p(x,t)\, \log x =  \frac{1}{2} 
\left\{ \log[4f(t)] + \frac{\rm d}{{\rm d} \nu}\log{\Gamma~\!\!(\nu+1)} \right\}\,\,,
\label{eq:mean_logx}\\
\frac{1}{2}\left< [x(t)]^2\right> &= \frac{1}{2}\,\int_0^{\infty}{\rm d}x\,p(x,t)\, x^2 = 2(\nu + 1)f(t)\,\,.
\label{eq:mean_x2}
\end{align}
These two averages also determine the mean internal energy of the system at the time $t$, 
$U(t) = \int_{0}^{\infty}{\rm d}x\,p(x,t)\,{\cal{V}}(x,t)= - g(t)\left< \log x(t)\right> + k(t)\left< [x(t)]^2\right>/2$.
Its increase from the beginning of the cycle,
\begin{equation}
\Delta U(t) = U(t) - U(0)\,\,,
\label{eq:internal_energy}
\end{equation}
and the mean work done from the beginning of the cycle up to the time $t$, $W_{\rm out}(t) \equiv W_{\rm out}(t,0)$, 
yield via the second law of thermodynamics the mean heat uptake during the time interval $[0,t]$:
\begin{equation}
Q(t) \equiv Q(t,0) = \Delta U(t) + W_{\rm out}(t)\,\,.
\label{eq:mean_heat}
\end{equation}

The total work done by the engine per cycle, $W_{\rm out} = W_{\rm out}(t_{\rm p})$, determines its average output power:
\begin{equation}
P_{\rm out} = \frac{W_{\rm out}}{\tper}\,\,.
\label{eq:POut}
\end{equation}
The efficiency of the engine is given by 
\begin{equation}
\eta = \frac{W_{\rm out}}{Q_{\rm in}}\,\,,
\label{eq:efficiency}
\end{equation}
where $Q_{\rm in} = \int_0^{\tper}{\rm d}t\,{{\rm d}Q(t,0)}/{{\rm d}t}\,\Theta\left[ {{\rm d}Q(t,0)}/{{\rm d}t} \right]$ stands for
the total heat transferred to the system from the reservoirs ($\Theta(x) = 1$ for $x>0$ and $0$ otherwise). In our setting $Q_{\rm in}$ equals to the heat uptake during the hot isotherm. In the rest of the paper we assume that $T_+>T_-$ and hence $Q_{\rm in} = Q(t_+^-,0) =  Q(t_+^+,0) = Q(t_+,0)$. The entropy of the system at the time $t$ is  $S(t) = - 
\int_0^{\infty}{\rm d}x\, p(x,t)\log\left[p(x,t)\right]$. Its increase from the beginning of the
cycle reads
\begin{equation}
S_{\rm s}(t) \equiv S(t) - S(0) = \frac{1}{2}\log\frac{f(t)}{f(0)}\,\,.
\label{eq:entropy_system}
\end{equation} 
The entropy transferred to the reservoirs during the time interval $[0,t]$ can be defined as
\begin{equation}
S_r(t) \equiv S_r(t,0) = -\int_0^t{\rm d}t'\,\frac{1}{T(t')}\frac{\rm d}{{\rm d}t'} Q(t')\,\,.
\label{eq:entropy_rezervoir}
\end{equation} 
Finally, the total entropy produced by the engine during the time interval $[0,t]$ 
is 
\begin{equation}
S_{\rm tot}(t) \equiv S_{\rm tot}(t,0) = S_{\rm s}(t,0) + S_r(t) \geq 0\,\,.
\label{eq:entropy_total}
\end{equation} 

\subsection{Fluctuations of work and power}
\label{sbs:fluctuations}
In order to investigate fluctuations of the above defined thermodynamic variables one needs to employ both the Green's function $R(x,t\,|\,x',t')$ and the probability density $p(x,t)$. These functions render all time-correlation functions of the underlying Markov process $x(t)$ and hence also all moments of the thermodynamic variables involved. For example the two-time correlation function 
\begin{multline}
\left< h[x(t)]\,f[x(t')] \right>_{\rm C} =\\= \left\{ 
  \begin{array}{l l}
    \,\int_{-\infty}^{\infty}{\rm d}x \int_{-\infty}^{\infty}{\rm d}x'\,
h(x)\,f(x') R(x,t\,|\,x',t')p(x',t')\,\,, & \quad t\geq t'\\
    \,\int_{-\infty}^{\infty}{\rm d}x \int_{-\infty}^{\infty}{\rm d}x'\,
f(x)\,h(x') R(x,t'\,|\,x',t)p(x',t)\,\,, & \quad t\leq t'
  \end{array} \right.
\label{eq:time_cor_fun_continuous}
\end{multline}
yields the the second raw moment of the random work done during the time window $[t,t']$ within the limit cycle, $w_{\rm out}(t,t') = \int_{t'}^t{\rm d}t''\,\partial {\cal V}\left[x(t''),t''\right]/\partial t''$, via the formula
\begin{equation}
\left<[w_{\rm out}(t,t')]^2\right> = \int_{t'}^t{{\rm d}t''}\, \int_{t'}^{t}{\rm d}t'''\, 
\left< \frac{{\partial}{\cal V}[x(t''),t'']}{{\partial}t''}\frac{{\partial}
{\cal V}[x(t'''),t''']}{{\partial}t'''} \right>_{\rm C}\,\,.
\label{eq:work_second_moment}
\end{equation}
Here, ${\partial}{\cal V}[x(t),t]/{\partial}t = - \log[x(t)]\,{\rm d}g(t)/{\rm d}t + 
[x(t)]^2/2\,{\rm d}k(t)/{\rm d}t$. Note that, 
due to the jumps in the driving, the functions ${\rm d}g(t)/{\rm d}t$ and ${\rm d}k(t)/{\rm d}t$ 
may also contain singular terms proportional to $\delta$-functions. For example 
${\rm d}k(t)/{\rm d}t = [k(t_+^+)-k(t_+^-)]\delta(t-t_+)+[k(t_{\rm p}^+)-k(t_{\rm p}^-)]\delta(t-t_{\rm p}) 
+ {\rm d}k(t)/{\rm d}t\left[\Theta(t_+-t) + \Theta(t_{\rm p}-t)\Theta(t-t_+)\right]$, where 
$\Theta(t)$ equals $1$ for $t > 0$ and $0$ otherwise. For an isothermal process the moment $\left<[w_{\rm out}(t,t')]^2\right>$ can be also obtained from the characteristic function of the work derived in \cite{Ryabov2013}.

The relative fluctuation of the output work,
\begin{equation}
\tilde{\sigma}_w(t,t') = \frac{\sqrt{\left<[w_{\rm out}(t,t')]^2\right>-[W_{\rm out}(t,t')]^2}}{|W_{\rm out}(t,t')|}\,\,,
\label{eq:work_out_variance}
\end{equation}
 determines the relative fluctuation of the output power,
\begin{equation}
\delta P_{\rm out} \equiv \tilde{\sigma}_w(t_{\rm p},0)\,\,,
\label{eq:power_out_variance}
\end{equation}
and hence, in a sense,  also the stability of the engine performance as we discuss in \se\ref{sec:discussion}. 

\section{Diagrams of the engine working cycle}
\label{sec:diagrams}

The mean work done by the engine per cycle can be written as $W_{\rm out}= W_{\rm g} + W_{\rm k}$. From \eq(\ref{eq:mean_work})
the two terms are identified:
\begin{align}
  \label{eq:W_g}
  W_{\rm g} &= \int_0^{t_{\rm p}} {\rm d}t\, \left<\log{x(t)}\right>\,\left[\frac{{\rm d}g(t)}{{\rm d}t}\right]\,\,, \\
  \label{eq:W_k}
  W_{\rm k} &= - \int_0^{t_{\rm p}} {\rm d}t\, \frac{1}{2} \left<[x(t)]^2\right> \frac{{\rm d}k(t)}{{\rm d}t}\,\,.
\end{align}
The first contribution equals the area enclosed by the parametric plot of the system response 
$\left<\log{x(t)}\right>$ versus the driving component $g(t)$, 
where the parameter $t$ runs from $0$ to $t_{\rm p}$. Similarly, the second contribution 
corresponds to the area enclosed by the parametric plot of the response $\left<[x(t)]^2\right>/2$ versus the driving component $- k(t)$. From now on we call the parametric plot corresponding 
to $W_{\rm g}$ ($W_{\rm k}$) as \emph{g-cycle} (\emph{k-cycle}). $W_{\rm g}$ and $W_{\rm k}$ 
are proportional to $(2\nu + 1)$ and to $(\nu + 1)$, respectively. Consequently, it is possible 
to tune the sign of the total output power $P_{\rm out} = (W_{\rm g} + W_{\rm k})/t_{\rm p}$ 
by changing the parameter $\nu>-1$. This effect is well visible in \fig{\ref{fig:efficiency_nu}}b), \se\ref{sec:discussion}, where the output power $P_{\rm out}$ as a function of the parameter $\nu$ is depicted.

Similar decomposition of work is
well known from classical thermodynamics. As an example, consider a periodically driven magnetic gas 
and let the driving possesses two components, the volume $V$ and the magnetic field
 (spatially homogeneous magnetic flux density) $B$.
The thermodynamic work done by the system per cycle then reads
$W_{\rm out} = \int_{V(0)}^{V(\tper)}p(V)\,{\rm d}V  - \int_{B(0)}^{B(\tper)}I(B)\,{\rm d}B$, where $p$ denotes the gas pressure and $I$ stands for the component of the total magnetic moment of the gas
parallel to the external magnetic field \cite{Callen2006}. The parametric plots corresponding to the terms $W_{\rm g}$ and $W_{\rm k}$ thus represent an analogy of the well known $p$--$V$ diagrams. In context of stochastic thermodynamics such plots were introduced in \cite{Chvosta2010}. Possible forms of the diagrams are discussed in \se\ref{sec:discussion} and illustrated in \figs\ref{fig:two_cycle} and \ref{fig:one_cycle} ibidem.

An important eye-guide in \figs\ref{fig:two_cycle} and \ref{fig:one_cycle} 
are the two equilibrium isotherms $\left< \log x(t)\right>_{\rm eq}$ and $\left<[x(t)]^2\right>_{\rm eq}/2$, 
where $\left<h(x)\right>_{\rm eq} = - \int_0^{\infty} {\rm d}x\, h(x) \exp\left[- {\cal V}(x,t)/T(t)\right]/Z(t)$,
$Z(t) = \Gamma(\nu + 1) 2^{\nu} [T(t)/k(t)]^{\nu+1}$. They correspond to the equilibrium values of the averages (\ref{eq:mean_logx}) and (\ref{eq:mean_x2}). The specific forms of  $\left< \log x(t)\right>_{\rm eq}$ 
and $\left<[x(t)]^2\right>_{\rm eq}/2$ are obtained if one substitutes
\begin{equation}  
  \label{eq:eq_f}
  f_{\rm eq}(t) = \frac{T(t)}{2k(t)}
\end{equation}
for the function $f(t)$ in \eqs(\ref{eq:mean_logx}) and (\ref{eq:mean_x2}), respectively. 
The farther a non-equilibrium isotherm is from the corresponding equilibrium one the more irreversible this branch is, cf. 
\fig\ref{fig:one_cycle}. Close to equilibrium the work probability density is reasonably approximated by 
a Gaussian distribution \cite{Speck2004, Speck2011, Nickelsen2011}. From Jarzynski equality \cite{Jarzynski1997}, valid 
during the isothermal branches, then follows  
$
T(t)\,\log \left[Z(t')/Z(t)\right] \approx - W_{\rm out}(t,t') - [W_{\rm out}(t,t')
\, \tilde{\sigma}_w(t,t')]^2/\left[2 T(t)\right]
$. 
We have used this formula for verification of the calculated functions 
(\ref{eq:mean_work}) and (\ref{eq:work_out_variance}). In the next \se\!\!, we specify
several forms of the component $k(t)$ of the driving with respect to the engine performance.

\section{External driving}
\label{sec:driving}

The quantities $P_{\rm out}$, $\delta P_{\rm out}$ and $\eta$ naturally determine  performance of a stochastic heat engine. The ideal engine should work not only at the largest possible output power (\ref{eq:POut}) with the smallest possible fluctuation (\ref{eq:power_out_variance}), but also with the largest possible efficiency (\ref{eq:efficiency}). However, such engine can not be constructed. 
For example, Carnot showed that the \emph{maximum possible efficiency} of a heat engine working between the temperatures $T_+$ and $T_-$, $T_+>T_-$, $\eta_{\rm C} = 1-T_{-}/T_{+}$,  is obtained in the 
reversible limit when the output power vanishes. Therefore one has to either maximize a single 
characteristic of the engine alone or settle with a compromise.

From the point of view of a possible experimental realisation of the engine, the potential exerted by an optical trap exhibits the desired shape only on a limited distance from its centre \cite{Cohen2005}. This means that the maximum size of the engine, which is proportional to the width of the distribution $f(t)$ is restricted. Another natural experimental restriction limits the maximum intensity of the laser which is proportional to $k(t)$.
 The arising optimization problem is hence to find a specific driving $k(t)$ which yields optimal performance of the engine for given maximum values of $f(t)$ and $k(t)$. 
However, it turns out that the corresponding mathematical problem is significantly simplified if one fixes the minimum $f(t)$ instead of the maximum $k(t)$. Physically, such modification is possible since the maximum $k(t)$ determines the minimum $f(t)$ via its equilibrium value (\ref{eq:eq_f}):
$\min\{f(t)\}_{t=0}^{t_{\rm p}} \geq \min \{f_{\rm eq}(t)\}_{t=0}^{t_{\rm p}}=\min\{T_+,T_-\}\bigg/\left[2\max\{k(t)\}_{t=0}^{t_{\rm p}}\right]$. In terms of power and efficiency the proposed optimization problem is solved in the next \sse.

\subsection{Maximum power and efficiency} 
\label{sec:max_power_driving}

Let the parameters $f_0 = f(0) = f(\tper)$, $f_1 = f(t_+)$, $t_{\pm}$, $T_{\pm}$ and $\nu$ are fixed. Let us now find the specific form of the function $k(t)$ which yields maximum power (\ref{eq:POut}) and efficiency for these parameters. From the optimization procedure it follows that the function $f(t)$ corresponding to the optimal driving is bounded by the values $f_0$, $f_1$ as demanded in the preceding paragraph. Later on we identify also ideal 
durations of the two isothermal branches, $t_{\pm}$, and verify general results concerning EMP obtained in \cite{Schmiedl2008, Esposito2009, Esposito2010b}.

The mean work (\ref{eq:mean_work}) 
represents a complicated non-local functional of $k(t)$ (see for example \cite{Then2008} where 
the procedure is performed for an isothermal process). Therefore, instead of finding the optimal 
protocol $k(t)$ directly, we adopt the procedure introduced by Schmiedl and Seifert \cite{Schmiedl2008} 
(see also \cite{Zhan-Chun2012}). The limit cycle (\ref{eq:cycle_dist}) 
corresponding to the (unknown) optimal protocol $k(t)$ is inevitably described by a 
certain function $f(t)$. This function can be obtained from $k(t)$ 
using \eq(\ref{eq:function_f}) and, similarly, the function $k(t)$ follows 
from the function $f(t)$ via the formula 
\begin{equation}
k(t) = - \frac{\dot{f}(t) - T(t)}{2f(t)}\,\,,
\label{eq:f_differential_eq}
\end{equation}
where $\dot{f}(t)\equiv{\rm d}f(t)/{\rm d}t$. \eqs(\ref{eq:function_f}) and (\ref{eq:f_differential_eq}) represent an one-to-one correspondence between the functions $k(t)$ and $f(t)$ and hence it does not matter 
if one first finds the optimal driving $k(t)$ or the optimal response $f(t)$. During the isothermal 
branches the work (\ref{eq:mean_work}) assumes the generic form
\begin{equation}
W_{\rm out}(t_{\rm f},t_{\rm i}) = W_{\rm ir}(t_{\rm f},t_{\rm i}) -[U(t_{\rm f}) - U(t_{\rm i})]
+  T S_{\rm s}(t_{\rm f},t_{\rm i})\,\,.
\label{eq:f_W_by_f}
\end{equation}
The first term in \eq(\ref{eq:f_W_by_f}) equals the \emph{irreversible work} \cite{Schmiedl2008}
\begin{equation}
W_{\rm ir}(t_{\rm f},t_{\rm i})=
- (\nu+1)\int_{t_{\rm i}}^{t_{\rm f}}{\rm d}t\,\frac{1}{f(t)}\left[ \dot{f}(t) \right]^2
=- T[S_{\rm tot}(t_{\rm f}) - S_{\rm tot}(t_{\rm i})] \leq 0\,\,,
\label{eq:irreversible_work}
\end{equation}
the increase of the internal energy is given by $[U(t_{\rm f}) - U(t_{\rm i})] = 2(\nu + 1)\left[k(t_{\rm f})f_{\rm f} - k(t_{\rm i})f_{\rm i} \right] - (2\nu + 1)/2\,T\log[f_{\rm f}\big/f_{\rm i}]$ and the third term proportional to the increase of the system entropy during the isothermal branch reads $T S_{\rm s}(t_{\rm f},t_{\rm i}) = T/2\log(f_{\rm f}/f_{\rm i})$. The work done during the first isotherm (the branch I in \fig\ref{fig:cycle_scheme}) is obtained after the substitution
$t_{\rm i} = 0$, $t_{\rm f} = t_+^-$, $f_{\rm i}=f(0) = f_0$, $f_{\rm f} = f(t_+) = f_1$ and 
$T = T_+$. The substitution $t_{\rm i} = t_+^+$, $t_{\rm f} = t_{\rm p}^-$, $f_{\rm i}= f(t_+) 
= f_1$, $f_{\rm f} = f(\tper) = f_0$ and $T = T_-$ yields the work done during the second isotherm (the branch III in \fig\ref{fig:cycle_scheme}).

Using the definition (\ref{eq:irreversible_work}), the total entropy produced per cycle, the work done 
per cycle and the efficiency can be rewritten as 
\begin{eqnarray}
S_{\rm tot} &=& S_{\rm tot}(\tper) = -\left[\frac{W_{\rm ir}(t_+,0)}{T_+} + \frac{W_{\rm ir}(t_{\rm p},t_+)}{T_-}\right]\,\,,
\label{eq:entropy_tot_min}\\
W_{\rm out} &=& (T_+ - T_-) S_{\rm s}(t_+,0) + [W_{\rm ir}(t_+,0) + W_{\rm ir}(t_{\rm p},t_+)]\,\,,
\label{eq:work_out_max}\\
\eta &=& \frac{W_{\rm out}}{Q(t_+,0)} = \left(1-\frac{T_-}{T_+}\right)\frac{W_{\rm out}}{W_{\rm out}+T_- S_{\rm tot}}\,\,,
\label{eq:efficiency_max}
\end{eqnarray}
respectively. The total output work is given by the sum of works done during the four branches of the cycle. In this sum the contributions from the internal energy cancels out. In the derivation of \eq(\ref{eq:efficiency_max}) we used the chain of identities: $W_{\rm out}+T_- S_{\rm tot} = (T_+ - T_-) [S_{\rm s}(t_+,0) + W_{\rm ir}(t_+,0)\big/T_+] = (T_+ - T_-)\big/T_+[W_{\rm out}(t_+^-,0) + U(t_+^-) - U(0)] = (T_+ - T_-)\big/T_+ Q(t_+,0)$.
Assume that the parameters $T_{\pm}$, $t_{\pm}$,
$f_0$ and $f_1$ are fixed. Moreover, let the function $f(t)$ maximize the irreversible 
work (\ref{eq:irreversible_work}) during the both isothermal branches. Then, \emph{for these 
fixed parameters}, \emph{this function} also \emph{minimizes the total entropy production }
(\ref{eq:entropy_tot_min}) and \emph{maximizes both the output work} (\ref{eq:work_out_max}) \emph{and 
the efficiency} (\ref{eq:efficiency_max}). The results are discussed and illustrated in \se\ref{sec:discussion},  \figs\ref{fig:efficiency_time} and \ref{fig:efficiency_nu}.  

The function $f(t)$ maximizing the integral (\ref{eq:irreversible_work}) solves
the Euler-Lagrange equation $[\dot f(t)]^2-2f(t)\,[\ddot f(t)] = 0$ 
with the boundary conditions $f(t_{\rm i})=f_{\rm i}$, $f(t_{\rm f})=f_{\rm f}$. Its solution 
for the isothermal branches I and III reads \cite{Schmiedl2007,Schmiedl2008, Then2008}
\begin{equation}
f(t) = \left\{ 
  \begin{array}{l l}
    \,\displaystyle f_0\left(1+A_1t\right)^2\,\,,\quad A_1 = \frac{1}{t_+}
    \left(\sqrt{\frac{f_1}{f_0}} - 1\right)\,\,, & \qquad t \in [0,t_+]\\
    \,\displaystyle f_1\left[1+A_2(t - t_+)\right]^2\,\,,\quad A_2 = 
    \frac{1}{t_-}\left(\sqrt{\frac{f_0}{f_1}} - 1\right)\,\,, & \qquad t \in [t_+,t_{\rm p}]
  \end{array} \right.
  \,\,.
\label{eq:f_max_power}
\end{equation} For this \emph{optimal response}, \eq(\ref{eq:f_differential_eq}) yields the 
following \emph{optimal protocol}:
\begin{equation}
k(t) = \left\{ 
  \begin{array}{l l}
    \,\displaystyle\frac{T_+}{2 f_0}\frac{1}{(1+A_1 t)^2} - \frac{A_1}{1+A_1 t}\,\,, & \qquad t \in [0,t_+^-]\\
    \\
    \,\displaystyle\frac{T_-}{2 f_1}\frac{1}{[1+A_2 (t-t_+)]^2} - \frac{A_2}{1+A_2 (t-t_+)}\,\,, & \qquad t \in [t_+^+,t_{\rm p}^-]
  \end{array} \right.
\,\,
\label{eq:driving_max_power}
\end{equation}
which represents
the solution of the problem proposed at the beginning of this \sse\!\! The assumptions we have imposed on the function $k(t)$ during the derivation of 
the optimal driving (\ref{eq:driving_max_power}) are the following: 1) $k(t)$ is 
continuous within the isothermal branches (otherwise the time derivative $\dot{f}(t)$ in 
\eq(\ref{eq:irreversible_work}) would not exist), 2) the optimal system response $f(t)$ 
corresponding to $k(t)$ is periodic and assumes the values $f_0$ and $f_1$ at the times 
$0$ and $t_+$, respectively.

The power corresponding to the optimal protocol 
(\ref{eq:driving_max_power}) reads
\begin{equation}
\widetilde{P}_{\rm out} =  \frac{T_+-T_-}{t_++t_-}S_{\rm s}(t_+,0) - A_{\rm ir}\frac{1}{t_+ t_-}\,\,,
\label{eq:power_max_power}
\end{equation}
where $A_{\rm ir} = 4(\nu+1)\left(\sqrt{f_1}-\sqrt{f_0}\right)^2$ stands for the irreversible 
action \cite{Schmiedl2008} which determines the heat uptake during the isothermal branches, 
$Q_{\pm} = \pm\, T_{\pm}S_{\rm s}(t_+,0) - A_{\rm ir}/t_{\pm}$, and thus also the total 
entropy produced per cycle, $S_{\rm tot} = - Q_+/T_+ - Q_-/T_- = [1\big/(t_+T_+) + 1\big/(t_- T_-)]A_{\rm ir}$. 
The authors of \cite{Schmiedl2008} already noted that, in the long time limit $t_{\pm} 
\rightarrow \infty$, the produced entropy converges to zero. This feature of the optimal driving stems from the fact that during its derivation one also minimizes the irreversible entropy production $S_{\rm tot}$ \cite{Zhan-Chun2012}. In this limit, the system is in thermal equilibrium during the \emph{whole cycle}. Indeed, from \eqs(\ref{eq:f_max_power}) and (\ref{eq:driving_max_power}) it follows that $\lim_{t_{\pm} \rightarrow \infty} f(t) = f_{\rm eq}(t)$ for all $t \in [0,t_{\rm p}]$. As was already noted by Sekimoto \cite{Sekimoto2000}, such long time behaviour can not be achieved by a driving without discontinuities at the times $t_+$ and $t_{\rm p}$ when the temperature changes suddenly. This observation is illustrated in \figs\ref{fig:one_cycle} and \fig\ref{fig:one_cycle_termodynamics} in \se\ref{sec:discussion}.

Further maximization of the power (\ref{eq:power_max_power}) with respect to the times $t_{\pm}$ 
leads to the optimal duration of the isothermal branches
\begin{equation}
\tilde{t}_+ = \tilde{t}_- = \frac{4 A_{\rm ir}}{(T_+-T_-) \,S_{\rm s}(t_+,0)}\,\,.
\label{eq:optimal_time}
\end{equation}
Without loss of generality we assume that the first reservoir is hot ($T_+>T_-$). 
Then the efficiency (\ref{eq:efficiency_max}) corresponding to the optimal 
protocol (\ref{eq:driving_max_power}) with the optimal period durations 
(\ref{eq:optimal_time}) can be written as
\begin{equation}
\eta_{\rm P} = \frac{2 \eta_{\rm C}}{4 - \eta_{\rm C}}\,\,,
\label{eq:optimal_efficiency}
\end{equation}
where $\eta_{\rm C} = 1 - T_-/T_+$ denotes the Carnot efficiency. 
Note that, contrary to the corresponding optimal power
$\widetilde{P}_{\rm out} = (T_+-T_-)^2 [S_{\rm s}(t_+,0)]^2/[16 A_{\rm ir}]$, $\eta_{\rm P}$ depends only on the reservoir temperatures $T_+$ and $T_-$. Under the assumption of small temperature difference ($\eta_{\rm C}$ small) one 
can perform Taylor expansion of (\ref{eq:optimal_efficiency}):
\begin{equation}
\eta_{\rm P} \approx \eta_{\rm CA} \approx \frac{\eta_{\rm C}}{2} + \frac{\eta_C^2}{8} + O\left(\eta_C^3\right)\,\,,
\label{eq:optimal_efficiency_taylor}
\end{equation}
where $\eta_{\rm CA} = 1 - \sqrt{T_-/T_+}$ stands for the Curzon-Ahlborn efficiency. The formulas (\ref{eq:optimal_efficiency}) and 
(\ref{eq:optimal_efficiency_taylor}) represent another example which verifies 
validity of general considerations of the works \cite{Schmiedl2008, Esposito2009, Esposito2010b} where the authors prove that the efficiency at maximum power (\ref{eq:optimal_efficiency}) is bounded as $\eta_{\rm C}/2 < \eta_{\rm P} < 
\eta_{\rm C}/(2-\eta_{\rm C})$ and that the expansion (\ref{eq:optimal_efficiency_taylor}) 
is universal for strong coupling models that possess a left-right symmetry.

\subsection{Minimum relative power fluctuation} 
Minimization of the relative power fluctuation (\ref{eq:power_out_variance}) 
is a task which is beyond the scope of the present paper (the corresponding functional is too complicated).  Using the physical 
intuition, one can guess that the driving which minimizes the work fluctuation 
$|W_{\rm tot}|\, \tilde{\sigma}_w(\tper,0)$ should perform all the work at the instants 
when the width of the particle distribution (\ref{eq:mean_x2}) is minimal. However, such driving would also perform 
zero amount of work $W_{\rm out}$. Let us now investigate a class of protocols where the work is performed only during the adiabatic branches. To this end we introduce the \emph{piecewise-constant driving}:
\begin{equation}
k(t) = \left\{ 
  \begin{array}{l l}
    \,\displaystyle r_1\,\,, & \qquad t \in [0,t_+^-]\\
    \,\displaystyle r_2\,\,, & \qquad t \in [t_+^+,t_{\rm p}^-]
  \end{array} \right.
\,\,.
\label{eq:driving_min_power_fluctuation}
\end{equation}
The protocol does not vary during the isotherms and thus these branches can be also referred to as \emph{isochoric}. 
Note that due to the jumps in $k(t)$ the resulting cycle can be performed quasi-statically.

\subsection{Fractional driving} 
Except for the protocols (\ref{eq:driving_max_power}) and 
(\ref{eq:driving_min_power_fluctuation}) we have 
examined also the \emph{fractional driving}:
\begin{equation}
k(t) = \left\{ 
  \begin{array}{l l}
    \,\displaystyle \frac{k_1}{1 + \gamma_1 t}\,\,, & \qquad t \in [0,t_+^-]\\
    \,\displaystyle \frac{k_2}{1 + \gamma_2 (t-t_+^+)}\,\,, & \qquad t \in [t_+^+,t_{\rm p}^-]
  \end{array} \right.
\,\,.
\label{eq:driving_fractional}
\end{equation}
Our results for this protocol coincide with those obtained in \cite{Ryabov2013}.  
Contrary to the protocols (\ref{eq:driving_max_power}) 
and (\ref{eq:driving_min_power_fluctuation}),
this driving may consist of two isotherms only 
(cf. with the driving used in the paper \cite{Chvosta2010}). In the figures 
we always take continuous $k(t)$, then the adiabatic branches vanish for
$\nu= - 1/2$ together with the logarithmic part of the potential. If the jumps in the driving are allowed, the engine corresponding to the fractional driving can be tuned so its performance
differs from that of the engine driven by the optimal protocol 
(\ref{eq:driving_max_power}) nearly negligibly. In the next \se we compare performances of the engines driven by the three protocols described above.

\section{Performance of the engine}
\label{sec:discussion}

\begin{figure}
	\centering
		\includegraphics[width=1.0\linewidth]{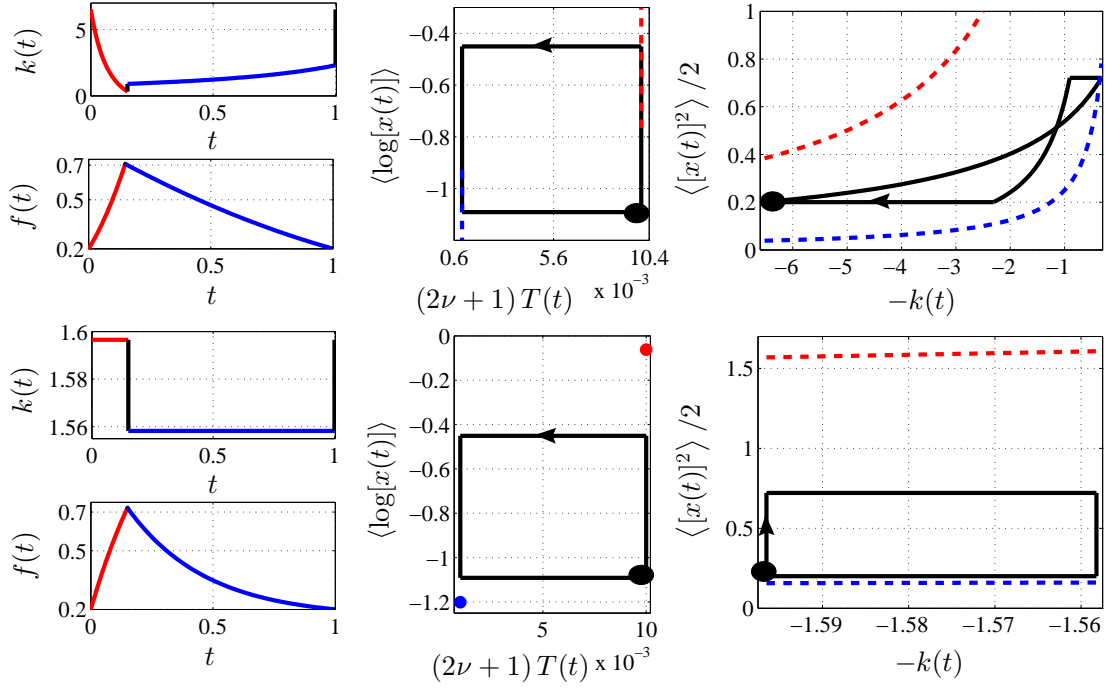}		
		\caption{ \emph{Upper panels} demonstrate the driving $k(t)$, the system response $f(t)$ 
		[\eq(\ref{eq:function_f})], the $g$-cycle (\ref{eq:W_g}) 
		formed by one counter-clockwise loop ($W_g<0$) and the $k$-cycle 
		(\ref{eq:W_k}) formed by one clockwise and one counter-clockwise loop ($W_k>0$) for the optimal
		 protocol (\ref{eq:driving_max_power}), respectively. \emph{Lower panels} show the same for the piecewise-constant driving 
		 (\ref{eq:driving_min_power_fluctuation}). The total work output 
		 for the both drivings is positive ($W_{\rm out}>0$). In the \emph{left panels} the red 
		 (blue) curve corresponds to the first (second) isotherm. The curves depicting the 
		 (infinitely fast) adiabatic branches are black. Note that the responses $f(t)$ for the two different protocols (in particular compare the ranges of used $k(t)$ 
     values) are quite similar. Red/blue dashed lines 
		 in the \emph{middle} and in the \emph{right panels} depict the hot/cold equilibrium isotherms.	
     In the both panels the black circles denote the initial points of the cycles,  
     directions of circulations are marked by the arrows.	During the isothermal branches 
     of the both $g$-cycles (and also during the isotherms of the $k$-cycle for the piecewise-constant driving) the driving is 
     constant and hence no work is produced. These branches thus represent an analogue of 
     isochores from the classical thermodynamics.	The piecewise-constant $k(t)$ for the piecewise-constant driving also results in the degeneration of the equilibrium isotherms 
corresponding to the g-cycle to points. For the both protocols we have used $t_+ = 0.15$, $t_- = 0.85$,
      $T_+ = 5$, $T_- = 0.5$, $f_0 = 0.2$, $f_1 = 0.7198$, $\nu = -0.499$ and thus the corresponding g-cycles coincide. Here and in all other figures the remaining parameters are calculated from the closure conditions on the driving and system response: $k(\tper) = k(0)$, $f(t_+^-) = f(t_+^+)$ and $f(\tper) = f(0)$.}		\label{fig:two_cycle}
\end{figure}

\begin{figure}
	\centering
		\includegraphics[width=1.0\linewidth]{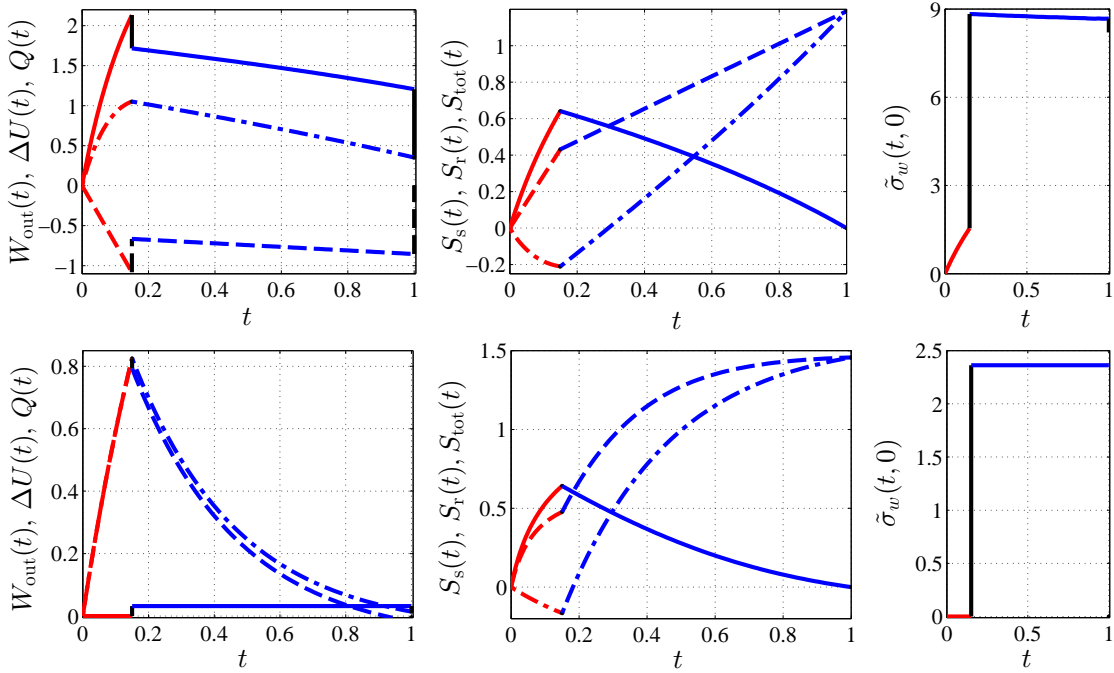}		
		\caption{Time-resolved thermodynamics of the engines depicted in 
		\fig\ref{fig:two_cycle}. The \emph{upper} (\emph{lower}) \emph{panels} in 
		\fig\ref{fig:two_cycle} correspond to the \emph{upper} (\emph{lower}) \emph{panels} herein. In all the panels the red (blue) curve stands for 
		the first (second) isotherm. The curves corresponding to the (infinitely fast) 
		adiabatic branches are black. 
		\emph{Left:} The increase of the internal energy of the system (broken lines) (\ref{eq:internal_energy}), the mean work done on the environment (solid lines), and the mean heat accepted by 
		the system (dot-dashed lines) (\ref{eq:mean_heat}). Note 
		that these curves verify the first law of thermodynamics. \emph{Middle:} The increase of the entropy of the 
		system (solid lines) (\ref{eq:entropy_system}), the entropy transferred to the reservoirs (dot-dashed lines) (\ref{eq:entropy_rezervoir}) and 
		the total entropy production of the engine (dashed lines) (\ref{eq:entropy_total}). The total entropy produced 
		per cycle,  $S_{\rm tot}(\tper)$, is always positive and equals the amount of the entropy 
		transferred to the reservoirs, $S_{\rm r}(\tper)$. \emph{Right:} The relative work fluctuation (\ref{eq:work_out_variance}). 
		The cycle driven by the optimal protocol (upper panels) produces larger work with larger 
		relative fluctuation than that corresponding to the piecewise-constant driving (lower panels). Note that the relative work fluctuation exhibits discontinuities during the adiabatic branches. Although this function can also decrease, the total relative fluctuation at the end of the cycle is always positive. More entropy is produced by the engine driven by the piecewise-constant protocol.}
    \label{fig:two_cycle_termodynamics}
\end{figure}

\begin{figure}
	\centering
		\includegraphics[width=1.0\linewidth]{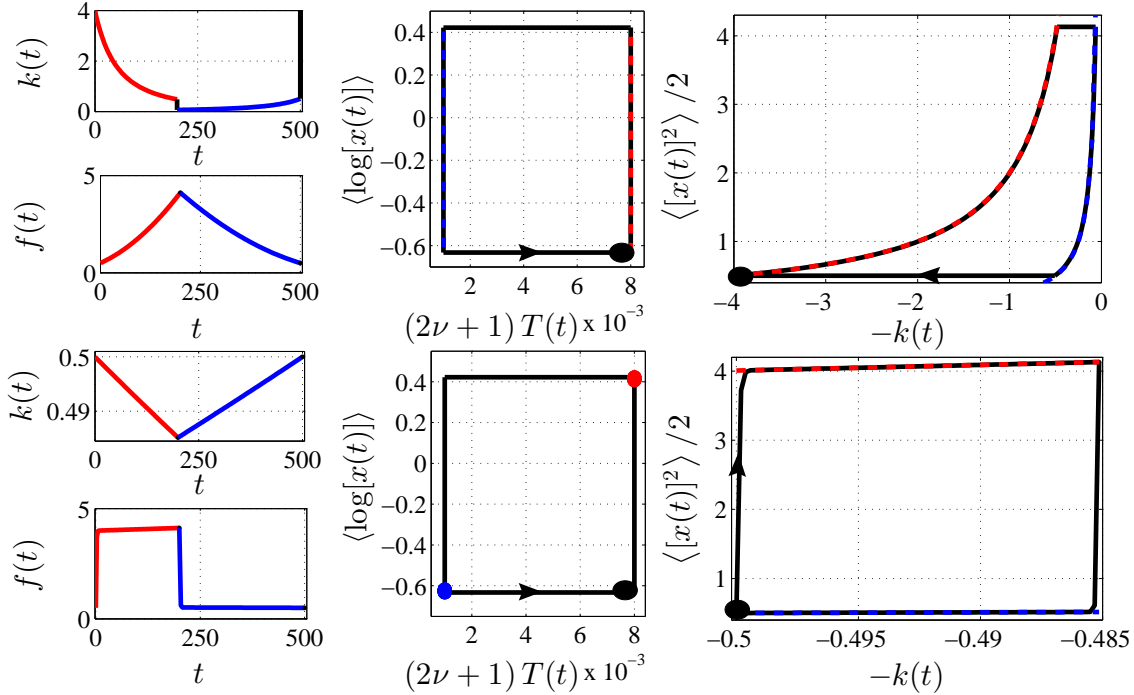}		
		\caption{The same quantities as in \fig{\ref{fig:two_cycle}}. \emph{Upper panels} correspond to the optimal protocol (\ref{eq:driving_max_power}) and \emph{lower panels} to the fractional driving 
		(\ref{eq:driving_fractional}). For the both cycles we have taken 
		very slow driving ($t_+ = 200$, $t_- = 300$). The cycle driven by the optimal protocol is nearly quasi-static and its efficiency $\eta \approx 0.868$ almost 
		reaches the Carnot's upper bound $\eta_{\rm C} = 0.875$. On the other hand, 
		during the cycle driven by the fractional driving the system is brought far from 
		equilibrium at the instants when the heat reservoirs are interchanged. During the 
		emerging relaxation processes a large amount of entropy is produced (cf. 
		\fig\ref{fig:one_cycle_termodynamics}) and the engine efficiency 
		$\eta \approx 0.352$ is far from $\eta_{\rm C}$. Also note that the corresponding system response 
		$f(t)$ changes after these time instants quite rapidly as compared to that for the optimal protocol. Due to the narrow interval of $k(t)$ values used for the fractional driving it may seem that the equilibrium isotherms in the corresponding $g$-cycle again degenerate to points. 
The range of $k(t)$ used for the optimal protocol is much larger. For the both protocols we have used $T_+ = 4$, $T_- = 0.5$, $f_0 = 0.5$, $f_1 = 4.1219$, 
		$\nu = -0.499$ and thus the corresponding g-cycles coincide. Note, however, that only the g-cycle for the optimal driving converges to the equilibrium isotherms.}		
		\label{fig:one_cycle}
\end{figure}

\begin{figure}
	\centering
		\includegraphics[width=1.0\linewidth]{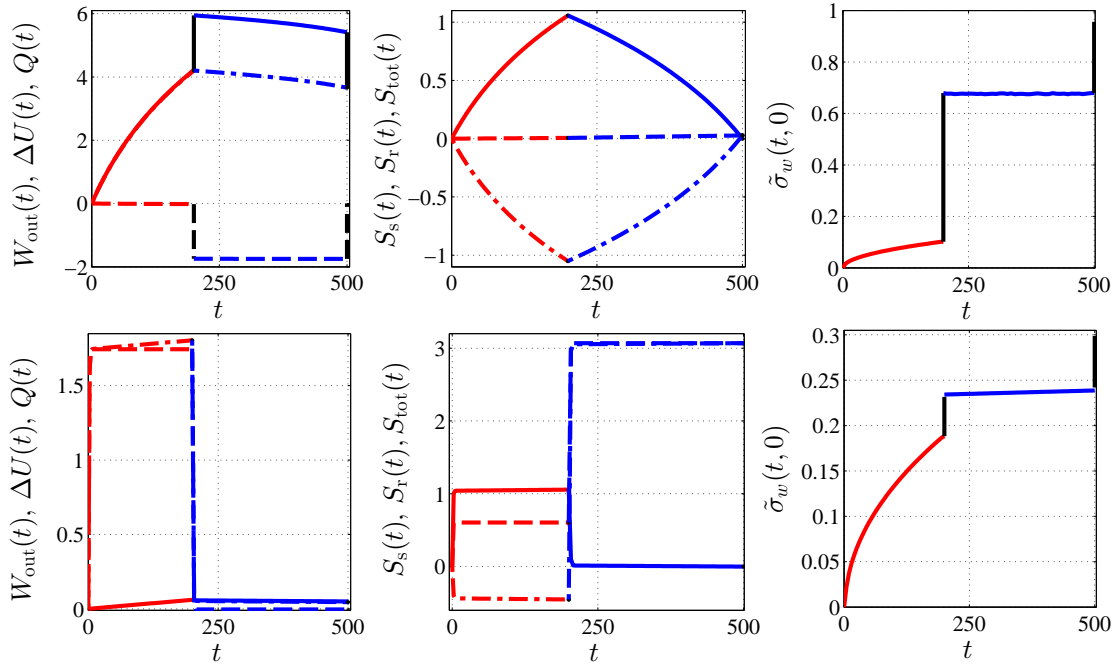}		
		\caption{Same quantities as in \fig\ref{fig:two_cycle_termodynamics} 
		for the engines discussed in \fig\ref{fig:one_cycle}. The engine driven 
		by the optimal protocol (upper panels) works during the whole cycle close to the quasi-static regime 
		(cf. \fig\ref{fig:one_cycle}). On the contrary, the cycle driven by the
		 fractional driving (lower panels) is brought far from equilibrium at the instants when the heat 
		 reservoirs are interchanged. This is reflected in large (small) output work	and small 
		 (large) entropy production corresponding to the optimal (fractional) protocol. Note that 
		 a considerable amount of the total entropy produced during the cycle with the fraction
     driving is created just after the two heat reservoirs are interchanged. The engine operating farther from equilibrium (fractional 
     driving) is favoured by the relative work fluctuation which is much smaller 
     than that corresponding to the optimal protocol.}	
     \label{fig:one_cycle_termodynamics}
\end{figure}

\begin{figure}
\centering
\includegraphics[width=1.0\linewidth]{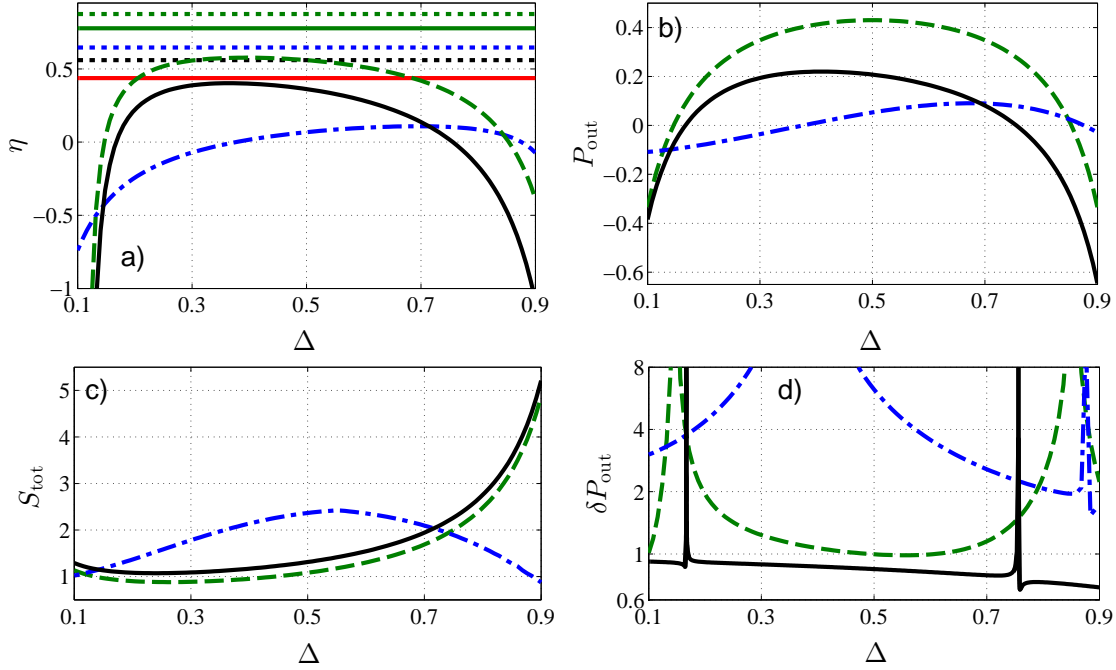}		
\caption{Performance of the engine as a function of the asymmetry 
parameter $\Delta = t_+/t_{\rm p}$ for the optimal
driving (\ref{eq:driving_max_power}) 
(green dashed lines), for the piecewise-constant driving 
(\ref{eq:driving_min_power_fluctuation}) 
(black solid lines) and for the fractional driving 
(\ref{eq:driving_fractional}) (blue dashed-dotted lines). 
The fixed $t_{\rm p}=2.2344$ 
is chosen in order to obtain the optimal time distribution 
(\ref{eq:optimal_time}) for $\Delta = 0.5$.
a) The efficiency (\ref{eq:efficiency}) of the engine. The upper (green), 
middle (blue) and lower (black) dotted horizontal lines stand for the Carnot efficiency $\eta_{\rm C}$, Curzon-Ahlborn efficiency 
$\eta_{\rm CA}$ and EMP $\eta_{\rm P}$
given by (\ref{eq:optimal_efficiency}), respectively. 
The upper green and lower red solid horizontal lines correspond
to the upper and lower bounds for EMP
$\eta_{\rm C}/(2-\eta_{\rm C})$ and $\eta_{\rm C}/2$. For $\Delta = 0.5$ the
efficiency for the optimal driving fulfils this limitation.
Note that the upper bound for EMP is
larger than the Curzon-Ahlborn efficiency.
b) The output power (\ref{eq:POut}). The maximal power
for the optimal driving is achieved for $\Delta = 0.5$.
c) The total entropy produced per cycle (\ref{eq:entropy_tot_min}).
d) The relative power fluctuation (\ref{eq:power_out_variance}) tends to $+\infty$ if the corresponding 
output power vanishes. We restricted the vertical limits in order to show 
the most interesting region of the figure in sufficient detail. The 
smallest power fluctuation is achieved by the piecewise-constant driving.
For all the protocols we took $T_+=4$, $T_-=0.5$, $f_0=0.5$, 
$\nu=-0.499$. Moreover, for the optimal and for the piecewise-constant driving we used $f_1=1.5$. For the fractional driving we took $k_1 = 1.8$ 
which results in a different $f_1$.
In agreement with the discussion in \se\ref{sec:max_power_driving}, for fixed $f_0$, $f_1$, $t_{\pm}$, $T_{\pm}$ and $\nu$ the optimal protocol yields maximum efficiency, output power and minimum amount of entropy. This is verified on
the piecewise-constant driving. For the fractional driving we 
have used different $f_1$ and thus the resulting efficiency and output power may exceed the values obtained for the optimal protocol.}
\label{fig:efficiency_time}
\end{figure}

\begin{figure}
\centering
\includegraphics[width=1.0\linewidth]{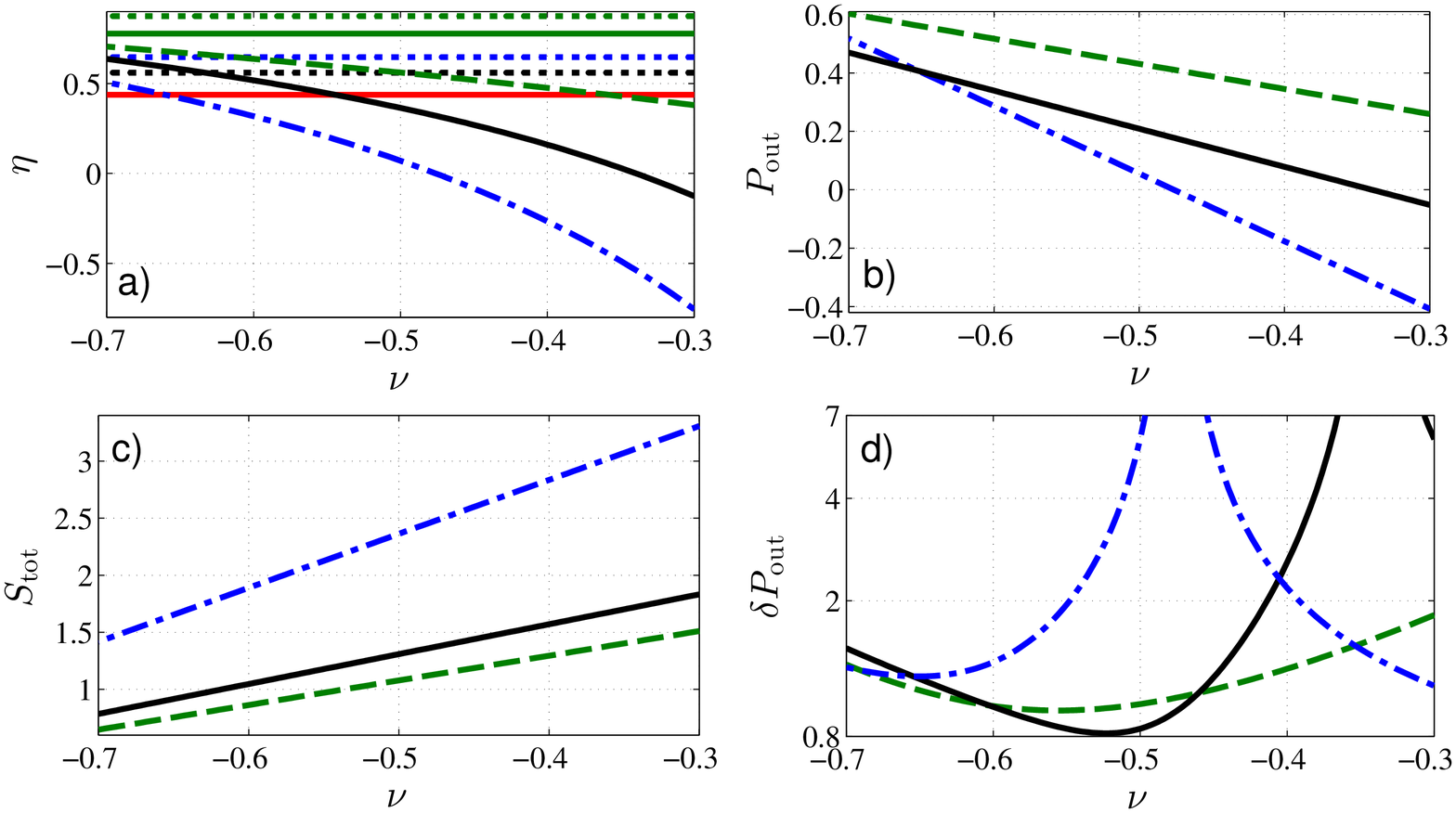}	
\caption{Performance of the engine as a function of the parameter
$\nu$ (\ref{eq:nu}). For $\nu> -0.5$ ($\nu<-0.5$) 
the logarithmic part of the potential (\ref{eq:potential})
is repulsive (attractive). The meaning of the individual curves and the remaining parameters are the 
same as in \fig\ref{fig:efficiency_time}. 
The efficiency, 
the output power and  the entropy produced per cycle are monotonic
functions of $\nu$. Note that EMP ($\nu=-0.499$) fulfils the restriction $\eta_{\rm C}/2< \eta_{\rm p} < \eta_{\rm C}/(2-\eta_{\rm C})$.
For the parameters taken, the optimal protocol yields the largest
efficiency, output power and the smallest entropy production 
from the three drivings. The relative power fluctuation both for the optimal protocol 
and for the piecewise-constant driving exhibit well pronounced minimum which is
deeper for the piecewise-constant driving.}
\label{fig:efficiency_nu}
\end{figure}


In the illustrations \fig\ref{fig:two_cycle} -- \fig\ref{fig:one_cycle_termodynamics} we assume that the protocols are
specified by the parameters $f_0 = f(0) = f(t_{\rm p})$, $f_1=f(t_+)$, $t_{\pm}$, $T_{\pm}$
and $\nu$. In all these figures the constant $f_0$ ($f_1$) represents the smallest (the largest) value of the function $f(t)$ during the cycle. This means that we compare the engines characterized by
the same minimum and maximum system size (volume).  For fixed $f_0$, $f_1$, $t_{\pm}$, $T_{\pm}$ and $\nu$ the parameters of the individual drivings are calculated using the closure conditions on the driving and system response: $k(\tper) = k(0)$, $f(t_+^-) = f(t_+^+)$, 
$f(t_{\rm p}) = f(0)$ and further, for the fractional driving, 
$k(t_+^-) = k(t_+^+)$, $k(\tper^-) = k(\tper)$. For arbitrary
reasonable parameters (positive $f_0$, $f_1$, $t_{\pm}$, $T_{\pm}$)
these formulas yield solution only for the optimal protocol. Therefore, in \figs\ref{fig:efficiency_time} and \ref{fig:efficiency_nu}, 
we were pushed to take different $f_1$ for the fractional driving than that used for other two protocols.

Two representative engine working cycles for the optimal protocol 
(\ref{eq:driving_max_power}) are depicted in 
\figs\ref{fig:two_cycle} and \ref{fig:one_cycle}. 
The corresponding time resolved thermodynamic quantities (\ref{eq:mean_work}), 
(\ref{eq:internal_energy}-\ref{eq:mean_heat}),
(\ref{eq:entropy_system}-\ref{eq:entropy_total}) 
and (\ref{eq:work_out_variance}) are 
depicted in \figs\ref{fig:two_cycle_termodynamics} and 
\ref{fig:one_cycle_termodynamics}. In \figs\ref{fig:two_cycle}, 
\ref{fig:two_cycle_termodynamics} and \ref{fig:one_cycle}, 
\ref{fig:one_cycle_termodynamics} we show examples of 
the engine working cycles together with the corresponding time resolved thermodynamic 
quantities for the fractional driving (\ref{eq:driving_fractional}) 
and for the piecewise-constant driving (\ref{eq:driving_min_power_fluctuation}), 
respectively. Contrary to the equilibrium situation (\ref{eq:eq_f}), when two isotherms at different temperatures never intersect, the simplest non-equilibrium cycle can be 
composed of only two isothermal branches. For the drivings used the $g$-cycle forms always a rectangle, while 
the $k$-cycle may exhibit two loops of a general shape. An example of such
two-loop cycle is depicted in \fig\ref{fig:two_cycle}. 

\figs\ref{fig:one_cycle} and \ref{fig:one_cycle_termodynamics} 
demonstrate the case of a very slow driving. The cycle driven by the optimal protocol is 
close to the quasi-static realization and its efficiency nearly attains the Carnot's upper 
bound $\eta_{\rm C} = 1-T_-/T_+$. In the limit of the infinitely slow driving 
($t_{\pm} \rightarrow \infty$), the sudden (adiabatic) changes of the optimal driving at 
the instants when the two reservoirs are interchanged guarantee that the equilibrium states 
before and after both adiabatic branches coincide \cite{Sekimoto2000}. On the other 
hand, during the cycle driven by the fractional driving, where the adiabatic changes of $k(t)$ are 
not considered, the system is brought far from equilibrium whenever the reservoirs are interchanged 
and the quasi-static limit does not exist ($S_{\rm tot}>0$ for any $t_{\rm p}$). 

The efficiency (\ref{eq:efficiency}), the output power (\ref{eq:POut}), the relative power 
fluctuation (\ref{eq:power_out_variance}) and the total entropy production 
(\ref{eq:entropy_total}) of engines driven by the protocols 
(\ref{eq:driving_max_power}), (\ref{eq:driving_min_power_fluctuation}) 
and (\ref{eq:driving_fractional}) are depicted in \figs\ref{fig:efficiency_time} 
and \ref{fig:efficiency_nu}. In \fig\ref{fig:efficiency_time} we study the 
dependence of these variables on the allocation of a given period $\tper$ between the two isotherms. In 
\fig\ref{fig:efficiency_nu} the engine performance as a function of the parameter $\nu$ is 
studied. It turns out that all the depicted quantities except the relative power fluctuation, which 
exhibits a well pronounced minimum, are monotonic functions of $\nu$. The curves corresponding
to the optimal protocol and to the piecewise-constant driving  verify that, for fixed parameters $f_0$, $f_1$, 
$t_{\pm}$, $T_{\pm}$ and $\nu$, the optimal protocol (\ref{eq:driving_max_power})
yields the maximum efficiency, output power and that it minimizes the entropy produced per cycle.
On the other hand, for the fractional driving we use different $f_1$ and hence the corresponding
efficiency and even the power output exceeds, for some parameters, the results obtained for the 
optimal protocol. For the parameters taken in \figs\ref{fig:efficiency_time} and
\ref{fig:efficiency_nu} the maximum power (\ref{eq:power_max_power}) 
is obtained for $\Delta = 0.5$ and for $\nu = -0.499$, respectively. In the both cases the efficiency at maximum power (\ref{eq:optimal_efficiency}) lies between the general bounds 
$\eta_{\rm C}/2$ and  
$\eta_{\rm C}/(2-\eta_{\rm C})$ which are depicted by the solid red and green 
horizontal lines.

The smallest relative power fluctuations observed in our illustrations are achieved by the 
piecewise-constant driving. Although this protocol does not actually minimize the power fluctuation, it shows limitations of the optimal driving (\ref{eq:driving_max_power}). 

\section{Conclusion and outlook}
\label{sec:conclusion}
We have investigated a stochastic heat engine based on an over-damped particle diffusing on the positive real axis in the externally driven time-periodic log-harmonic potential (\ref{eq:potential}). The periodic driving (\ref{eq:cycle_energies}) was composed of two isothermal and two adiabatic branches. We have found the Green's function solving the Chapman-Kolmogorov equation for the periodic driving (\ref{eq:green_function_cycle}) and also the periodic state of the engine during its working cycle (\ref{eq:cycle_dist}). These two functions allowed us to investigate performance of the engine in terms of mean values of output work, power end efficiency (\sse\ref{sbs:mean_values}) and also in terms of power fluctuations  (\sse\ref{sbs:fluctuations}). Namely, we have derived a specific protocol which, for certain fixed parameters, maximizes both the output power and the efficiency (\ref{eq:driving_max_power}). Using this protocol we have verified recent universal results regarding the efficiency at maximum power \cite{Schmiedl2008, Esposito2009, Esposito2010b} within our specific setting. Moreover, we have designed the protocol (\ref{eq:driving_min_power_fluctuation}) which shows that the driving favoured by the maximum power is not optimal with respect to power fluctuations (see \figs\ref{fig:two_cycle_termodynamics}, \ref{fig:one_cycle_termodynamics}, \ref{fig:efficiency_time} and \ref{fig:efficiency_nu}). 
This fact could disadvantage the power maximizing protocol in applications where sharp non-fluctuating values of power are needed. Such applications would rather utilize a protocol which minimizes the power fluctuation. To the best of our knowledge, the corresponding driving was not studied yet. The mathematical problem in question is quite complicated. It would be interesting, however, to investigate such protocol at least  numerically as it was done by Then and Engel \cite{Then2008} for the power maximizing protocol.

From macroscopic world we know that the power fluctuations towards 
large values can be handled by enlarging the capacity of the system where the power is delivered. 
On the other hand, in order to balance the fluctuations towards small values a standby power supply must be used. An example is the discussion about wind power plants which cause 
current fluctuations in transmission-grids \cite{Eriksen2005}. The natural Brownian motors 
\cite{Jung1990,  Hanggi2005, Hanggi2009, Astumian2002} are exposed to large fluctuations of 
the environment and hence they are quite adapted, indeed. Nevertheless, it is an interesting question
whether the principle of minimal power fluctuations, the principle of maximum output power, or
some other principle eventually applies in this field. In any case, these notions
should be considered during the design of artificial Brownian motors \cite{Hanggi2009}.

\section*{Acknowledgements}
The author would like to thank the anonymous referee for his thorough reading of the manuscript and many helpful suggestions and to Petr Chvosta for his kind guidance and help during our long cooperation. Support of this work by the Grant Agency of the Charles University (grant No. 301311) and by the project SVV - 2013 - 263 301 of the Charles University in Prague is gratefully acknowledged.
\section*{References}
\bibliographystyle{plain}
\bibliography{MotorReferences}


\end{document}